\pdfoutput=1

\documentclass[final,1p,times]{elsarticle} 
\usepackage{graphicx}
\usepackage{amssymb}
\usepackage{amsthm}
\usepackage{lineno}

\usepackage{textcomp}
\usepackage{ifpdf}
\usepackage{xspace}
\usepackage{float}
\usepackage{natbib}

\usepackage[colorlinks=true]{hyperref}

\newcommand{\driftvolume}{90\,m$^3$\xspace}
\newcommand{\gas}{\mbox{Ne-CO$_2$-N$_2$}\xspace}
\newcommand{\mixture}{\mbox{[85.72\%-9.52\%-4.76\%]}\xspace}
\newcommand{\nchannels}{557\hspace{0.25ex}568\xspace}
\newcommand{\area}{\mbox{32.5\,m$^2$}\xspace}

\journal{Nuclear Physics A} 

\begin{document}

\begin{frontmatter} 


\title{Commissioning and Calibration of the ALICE TPC}

\author{J.~Wiechula, for the ALICE TPC Collaboration}

\address{Physikalisches Institut, Universit\"at Heidelberg, Philosophenweg 12, 69120 Heidelberg, Germany}

\begin{abstract} 
ALICE (A Large Ion Collider Experiment) is the dedicated heavy ion experiment at the Large Hadron
Collider at CERN. The main tracking device of ALICE is a large volume TPC.

The milestones of the TPC commissioning as well as the current status
of the detector calibration are presented. The
obtained resolutions in transverse momentum, position as well as in specific energy loss (dE/dx) are presented and results from noise and electron drift velocity measurements are addressed.
\end{abstract} 

\end{frontmatter} 

\section{Introduction}
The ALICE TPC~\cite{TDR:tpc} is a cylindrical \driftvolume volume Time Projection Chamber of about 5\,m in length as well as in diameter.

A central electrode (CE) divides the drift volume into two readout sides. The nominal drift field of 400\,V/cm is achieved by applying 100\,kV to the CE.
The endcaps of the TPC are divided into 18 sectors. 
Each sector is instrumented with an inner (IROC) and outer (OROC) multiwire proportional readout chamber with cathode pad readout.
The 72 readout chambers have a total active area of \area, segmented into \nchannels readout pads. To account for the decreasing track density towards the outer radius, three different pad sizes are used.

The choice of the gas mixture \gas\mixture\cite{Garabatos2004} 
results in a stringent requirement for the temperature
stability and homogeneity: over the full volume of the TPC the temperature must be kept within 0.1\,\textcelsius\,\cite{Wiechula2005}.
This results from the non-saturated drift velocity and the required overall systematic error on the position resolution of less than 100\,\textmu m.

\section{Commissioning}
After the assembly of the TPC was finished in 2006, a first commissioning was carried out. 
Because a limited amount of services was available, only two sectors were operated at a time.
After the successful test of all readout chambers and the front-end electronics, the TPC was transported to the experimental area underground in the beginning of 2007.
By the end of 2007 first data taking with cosmic particle triggers took place.

In 2008 extensive data taking over a period of more than six months was carried out. 
In this stage of the commissioning all components of the TPC showed stable operation.

The recirculating gas system is in operation since the first commissioning phase in 2006. Only small variations in the gas mixture on a time scale of several days are observed.

The readout chambers were operated at nominal voltages without high-current events (trips).

The complex cooling system was commissioned during 2008. About 60 temperature-adjustable cooling loops combined with around 500 temperature sensors located around and inside the TPC allow for a precise measurement and control of the temperature distribution of the device. By adjusting the temperatures of the individual cooling loops in an iterative procedure, the required temperature stability and homogeneity of less than 0.1\textcelsius{} was met.

\section{Calibration and performance}
\begin{figure}
  \centering
  \includegraphics[width=0.49\textwidth]{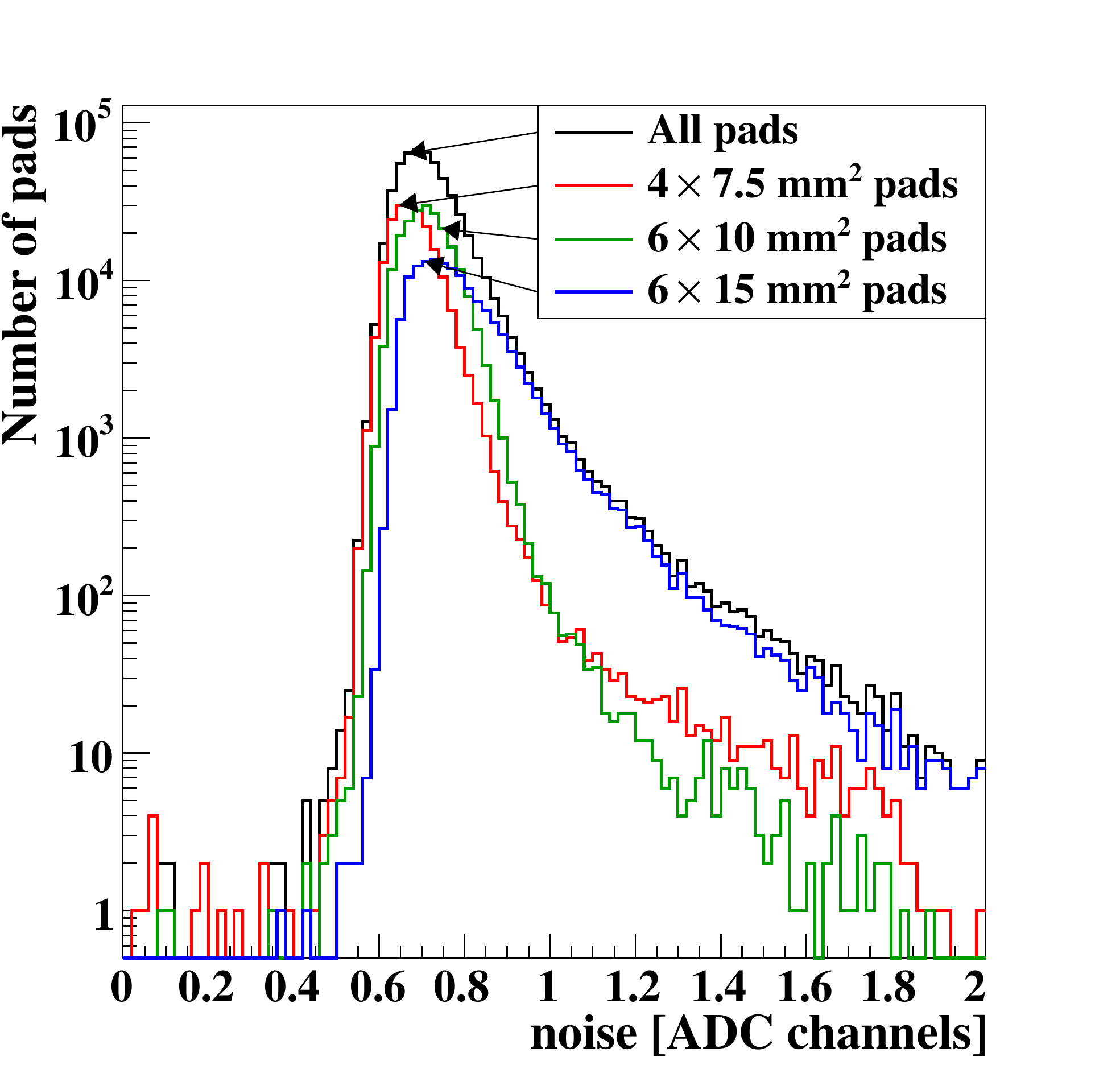}
  \caption{Noise distribution, both for all readout channels and the different pad sizes separately. 1\,ADC channel corresponds to an equivalent noise charge of about 1000\,e$^-$.}
  \label{fig:noise}
\end{figure}

Fig.~\ref{fig:noise} shows the measured noise within the TPC, both for all readout channels and the 
different pad sizes separately. The mean noise achieved within the detector, about 700\,e$^-$ equivalent noise charge, is better than the value specified in the technical design resport~\cite{TDR:tpc} (1000\,e$^-$). 

Due to the large raw data volume ($\sim$700\,MB/event) zero suppression is already performed at the level of the front-end electronics. Stable noise conditions result in a zero suppressed raw data volume less than 70\,kB/event for empty events, allowing for high data rates of up to 1\,kHz in pp and a few 100\,Hz in central PbPb collisions at LHC energies.

\begin{figure}
  \centering
  \includegraphics[width=0.49\textwidth]{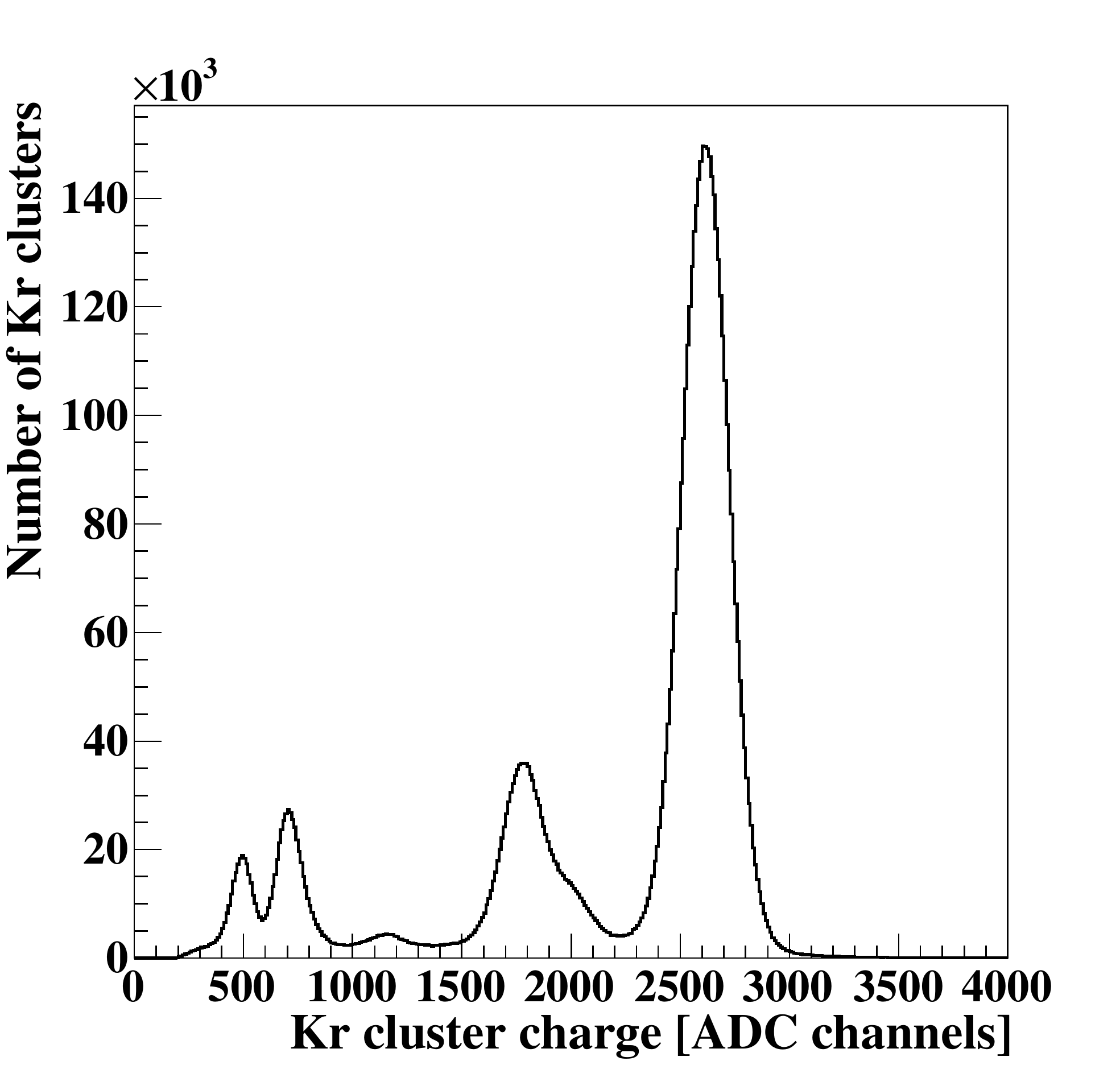}
  \caption{Integrated Kr spectrum of all OROCs.}
  \label{fig:krypton}
\end{figure}

An absolute gain calibration of each individual readout channel is obtained by releasing 
radio-active $^{83}_{36}$Kr into the TPC gas. 
For each channel the decay spectrum is acquired. 
Calibration constants are obtained by fitting the main (41.6\,keV) peak of the spectrum with a Gaussian function. 
The resulting 0.2\% error on the mean value is well below the required 1.5\% on the single pad level. 
Fig.~\ref{fig:krypton} shows the integrated pulse height spectrum of all OROCs.
The resolution of the main peak integrated over the IROCs and OROCs is 4.0\% and 4.2\%, respectively. 
During a period of two days 11.3\,$\times$\,10$^6$ events were recorded, each containing around 80 clusters from Kr decays.

\begin{figure}
  \centering
  \includegraphics[width=0.49\textwidth]{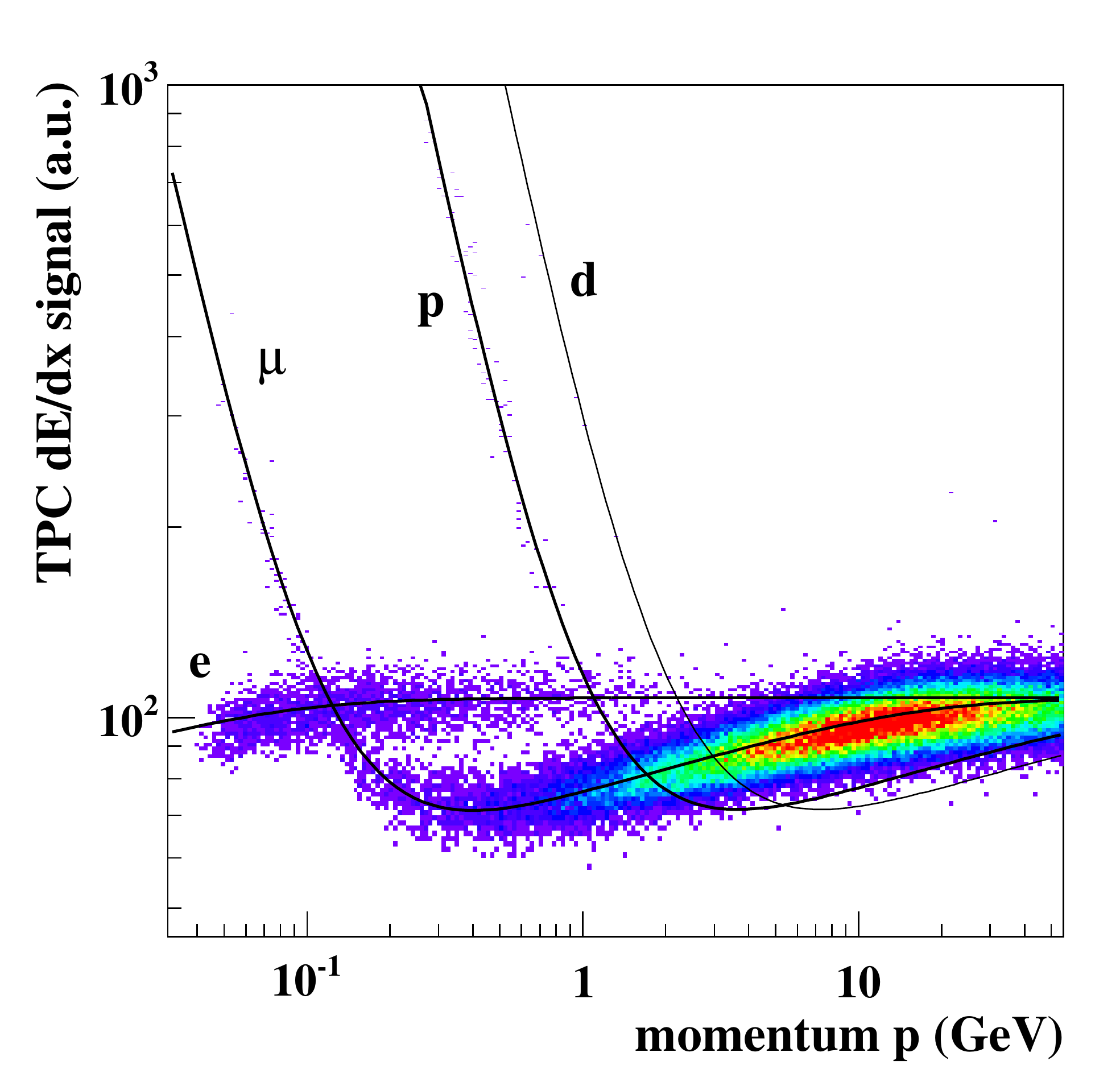}
  \caption{Specific energy loss of particles in the TPC measured with a trigger on cosmic particles.}
  \label{fig:dEdx}
\end{figure}

A study of the d$E$/d$x$ resolution of the TPC was carried out using $\sim$7$\times$10$^5$ tracks recorded during the data taking in 2008 with a cosmic particle trigger. Fig.~\ref{fig:dEdx} shows the measured d$E$/d$x$ as a function of the particle momentum.
The obtained resolution of 5.7\% is already close to the design value of 5.5\%.

\begin{figure}
  \centering
  \includegraphics[width=0.48\textwidth]{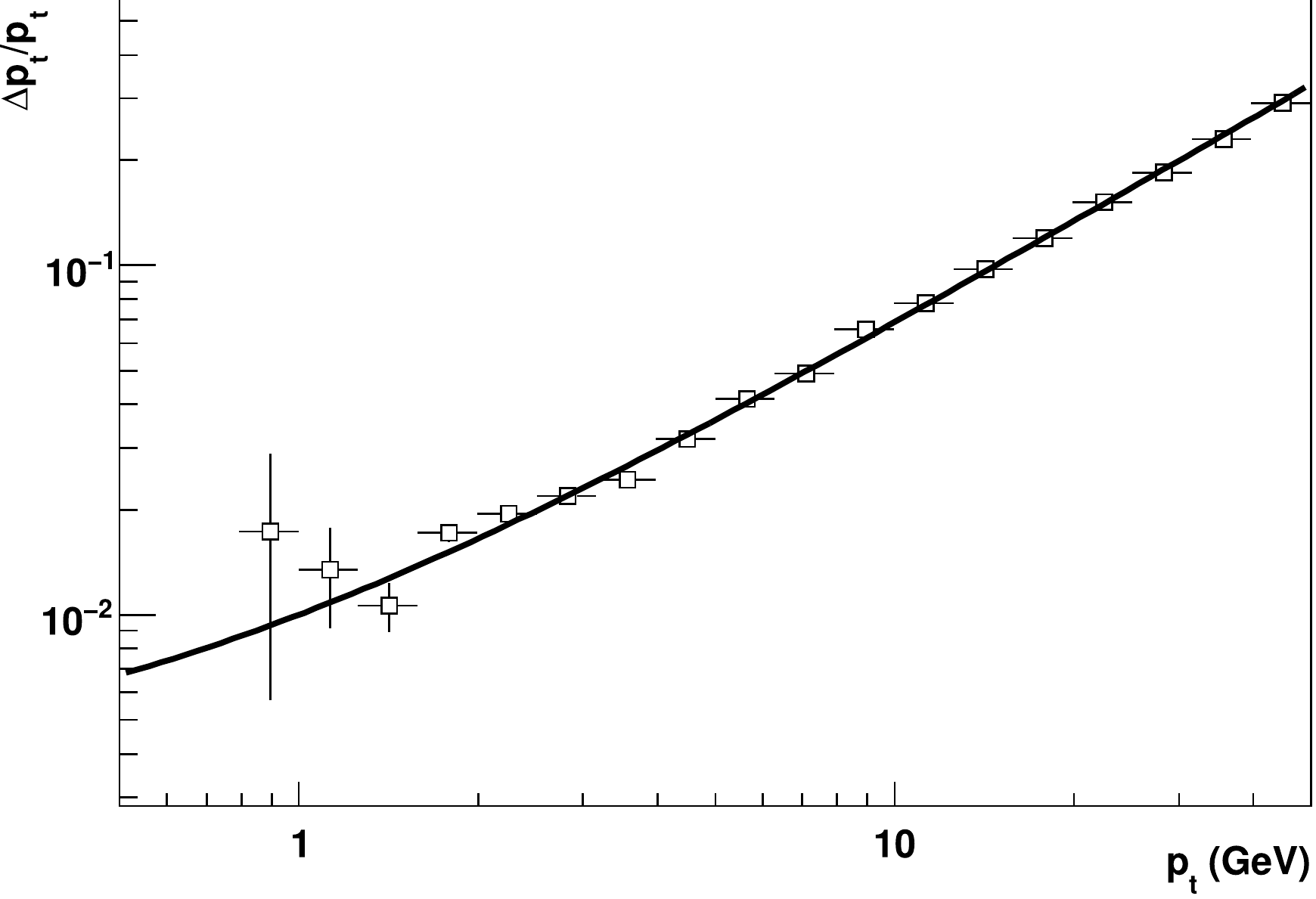}
  \caption{Relative momentum resolution.}
  \label{fig:pt}
\end{figure}

The transverse momentum (p$_t$) resolution is determined by using tracks traversing the complete TPC. 
Those are reconstructed in two parts due to the optimization of the reconstruction code for the collider geometry.
The relative p$_t$ resolution of the upper and lower half track (divided by $\sqrt{2}$) as a function of p$_t$, obtained from a study using about 5$\times$10$^5$ tracks from cosmic particles is shown in fig.~\ref{fig:pt}. The measured resolution at 10\,GeV of 6.5\% is at the current stage of calibration slightly above the design value (4.5\%).

The measured space point resolution in r$\varphi$-direction is in agreement with simulations. For low pad inclination angles (high momentum tracks) it was determined to be 300-800\,\textmu m, depending on the drift length (diffusion).

To obtain the required drift velocity resolution of 10$^{-4}$, precision measurements will be performed on a minutely basis. Different algorithms using physics tracks as well as information from the built-in laser system are implemented. The obtained resolution is on the order 2$\times$10$^{-5}$. In between the measurements a correction for the gas density effect is sufficient by following the temperature and pressure changes within the TPC.

\section{Summary}
During 2008 the main tracking device of the ALICE experiment at the CERN LHC, a large volume Time Projection Chamber, was fully commissioned. Extensive data taking took place over a period of more than six months. Already at the current, preliminary state of calibration the detector performs very close to specifications.

\bibliography{general}

\begin{thebibliography}{1}

\bibitem{TDR:tpc}
ALICE{ Collaboration},
\newblock \href{https://edms.cern.ch/document/398930/1}{{\em Technical Design
  Report of the Time Projection Chamber}},
\newblock CERN/LHCC 2000-001, 2000.

\bibitem{Garabatos2004}
C.~Garabatos,
\newblock
  \href{http://www.sciencedirect.com/science/article/B6TJM-4D2X8RC-M/2/fce5ba1%
18ed40946e6f6488f6e48954b}{{\em The ALICE TPC}},
\newblock Nucl. Instr. Meth. {\bf A535}, 197 (2004),
\newblock \href {http://dx.doi.org/10.1016/j.nima.2004.07.127}
  {\path{doi:10.1016/j.nima.2004.07.127}}.

\bibitem{Wiechula2005}
J.~Wiechula {\em et~al.},
\newblock
  \href{http://www.sciencedirect.com/science/article/B6TJM-4GFV296-3/2/1c0fc05%
c48c1d3c76a7b65a0f1b2ec1d}{{\em High-precision measurement of the electron
  drift velocity in Ne-CO2}},
\newblock Nucl. Instr. Meth. {\bf A548}, 582 (2005),
\newblock \href {http://dx.doi.org/10.1016/j.nima.2005.05.031}
  {\path{doi:10.1016/j.nima.2005.05.031}}.

\end{thebibliography}

\end{document}